\documentclass[reprint,
superscriptaddress,
amsmath,amssymb,
aps, physrev,
prb,
floatfix,
]{revtex4-2}

\usepackage{color}
\usepackage{graphicx}
\usepackage{dcolumn}
\usepackage{bm}

\usepackage{upgreek}
\usepackage{bm}
\usepackage{physics}
\usepackage{comment}

\begin{document}

\preprint{APS/123-QED}

\title{Significant modulation of acoustoelectric current associated with charge density wave transitions}

\author{Natsumi Nikaido}
\affiliation{Department of Physics, Graduate School of Science, The University of Osaka, Osaka 560-0043, Japan}

\author{Takuya Kawada}
\email[]{takuyakawada@g.ecc.u-tokyo.ac.jp}
\affiliation{Department of Physics, Graduate School of Science, The University of Osaka, Osaka 560-0043, Japan}
\affiliation{Department of Basic Science, The University of Tokyo, Tokyo 153-8902, Japan}

\author{Koji Fujiwara}
\affiliation{Department of Physics, Graduate School of Science, The University of Osaka, Osaka 560-0043, Japan}

\author{Jihoon Park}
\affiliation{Department of Physics, Graduate School of Science, The University of Osaka, Osaka 560-0043, Japan}

\author{Nan Jiang}
\affiliation{Department of Physics, Graduate School of Science, The University of Osaka, Osaka 560-0043, Japan}
\affiliation{Center for Spintronics Research Network,  The University of Osaka, Osaka 560-8531, Japan}
\affiliation{Institute for Open and Transdisciplinary Research Initiatives, The University of Osaka, Osaka 565-0871, Japan}

\author{Kouta Kondou}
\affiliation{Institute for Open and Transdisciplinary Research Initiatives, The University of Osaka, Osaka 565-0871, Japan}

\author{Shintaro Takada}
\affiliation{Department of Physics, Graduate School of Science, The University of Osaka, Osaka 560-0043, Japan}
\affiliation{Institute for Open and Transdisciplinary Research Initiatives, The University of Osaka, Osaka 565-0871, Japan}
\affiliation{Center for Quantum Information and Quantum Biology, The University of Osaka, Osaka 560-0043, Japan}

\author{Yasuhiro Niimi}
\affiliation{Department of Physics, Graduate School of Science, The University of Osaka, Osaka 560-0043, Japan}
\affiliation{Center for Spintronics Research Network,  The University of Osaka, Osaka 560-8531, Japan}
\affiliation{Institute for Open and Transdisciplinary Research Initiatives, The University of Osaka, Osaka 565-0871, Japan}

\date{\today}

\begin{abstract}
We studied acoustoelectric (AE) currents in materials that undergo charge density wave (CDW) transitions, induced by a surface acoustic wave (SAW) on a piezoelectric substrate. 
The polarity and magnitude of the AE current in NbSe$_3$ and 2H-TaSe$_2$ were modulated due to their CDW transitions. 
We also found that the sign of the AE current depends on the SAW propagation direction with respect to the crystalline axis of the substrate.
A phenomenological model assuming strain-modified conductivity can qualitatively account for the significant modulation of the AE current associated with the CDW transition, as well as the SAW propagation orientation dependence. The present results offer a powerful probe for exploring SAW-electron interactions in van der Waals materials, thereby highlighting their potential for advancing the emerging field of straintronics.
\end{abstract}

\maketitle




\section{Introduction}
Surface acoustic waves (SAWs) are vibrational modes localized at solid surfaces. Coherent excitation of SAWs can be achieved by applying an rf signal to an interdigital transducer (IDT) patterned on a piezoelectric substrate~\cite{white1965}. Their relatively low phase velocities, typically about five orders of magnitude smaller than the speed of light, enable device dimensions in the micro- to nano-scale range. These characteristics offer significant advantages for applications, and SAWs have consequently been used in diverse fields ranging from communication technology to biology~\cite{delsing2019}. In addition to their technological applications, SAWs play important roles in fundamental physics by providing a versatile means to probe and control electronic states in various materials. Both the charge and spin degrees of freedom of electrons can couple to SAWs, enabling, for example, the investigation of quantum Hall states~\cite{wixforth1986,esslinger1992,falko1993}, the transport of electrons~\cite{hermelin2011} and excitons~\cite{peng2022}, coupling to superconducting qubits~\cite{martin2014,manenti2017}, and the excitation of magnetic resonance~\cite{weiler2011,kobayashi2017prl}.

While conventional studies on SAW–electron coupling have primarily employed thin films grown directly on piezoelectric substrates, recent advances in van der Waals (vdW) material science have significantly broadened the range of accessible material systems. VdW crystals exhibit high crystallinity and host various electronic phases which can persist even in atomically thin flakes.
Furthermore, different vdW materials can be assembled into heterostructures with exceptional tunability in material composition, stacking configuration, thickness, and interlayer twist angle~\cite{geim2013,cao2018}, providing an ideal platform for exploring novel SAW–electron interactions~\cite{yokoi2020,peng2022,zhao2022,lyons2023}.
However, the lateral dimensions of exfoliated vdW flakes are typically several orders of magnitude smaller than the region where the SAW propagates. Consequently, only a small fraction of the SAW interacts with the electrons in the vdW flakes. As a result, conventional SAW transmission measurements, which are commonly used to evaluate SAW–electron interactions, become challenging in such systems~\cite{fang2023}.

From this perspective, the acoustoelectric (AE) current~\cite{parmenter1953,weinreich1957} provides an alternative probe of SAW–electron coupling in microscale vdW samples. In conducting materials, SAWs generate a dc electric current i.e., the AE current. The AE current exhibits several characteristic features. Its polarity is determined by the direction of the SAW wave vector (forward or backward)~\cite{rotter1998,miseikis2012} and the carrier type (electrons or holes)~\cite{bandhu2016,sun2024}. Its magnitude is linearly proportional to the SAW power~\cite{wang1962,rotter1998,miseikis2012}. Although experimental studies on the AE effect have mainly focused on graphene~\cite{miseikis2012,poole2017} and a limited number of vdW semiconductors~\cite{preciado2015,zheng2018}, AE current measurements are, in principle, applicable to metals and semiconductors. Thus, they offer a versatile means of probing SAW–electron interactions in a number of vdW materials.

Among various vdW systems, here we focus on materials that undergo a charge-density-wave (CDW) transition. A CDW state is characterized by a periodic modulation of the electron density and is typically realized in low-dimensional conductors with strong electron–lattice interactions~\cite{gruner2018}. 
While CDW physics has a long-standing history, recent advancement of vdW materials science combined with matured nanofabrication technology has brought renewed interest in the field~\cite{goli2012,yu2015,liu2016,fujiwara2021,zheng2023,ghosh2025,taheri2026}. 
Previous studies have shown that static strain can modulate the transport properties of CDW materials~\cite{lear1984,tseng1995} and that dynamic (ac) strain can couple to sliding CDWs~\cite{nikitin2021}. Several theoretical works have proposed that SAW-induced strain can dynamically couple to CDW states~\cite{funami2023,mori2023}. Moreover, recent experiments have demonstrated CDW dynamics driven by SAW-induced strain~\cite{fujiwara2025prl}, thereby motivating a systematic investigation of the AE current in CDW materials.

In this work, we have investigated the AE current in materials exhibiting CDW transitions. We observe a polarity reversal and an enhancement of the AE current in the CDW states of NbSe$_3$ and 2H-TaSe$_2$ exfoliated on piezoelectric LiNbO$_3$ substrates. We also find that the sign of the AE current depends on the SAW propagation direction relative to the crystallographic axes of the LiNbO$_3$ substrate, indicating that both the electric and strain fields associated with the SAW contribute to the AE current generation. To account for these observations, we model the AE current as arising from the rectification of the SAW-induced electric field via strain-induced modulation of the conductivity, which qualitatively reproduces the experimental results.

\section{Experimental Methods}

\begin{figure}[t]
	\begin{minipage}{1.0\hsize}
		\centering
		\includegraphics[scale=0.105]{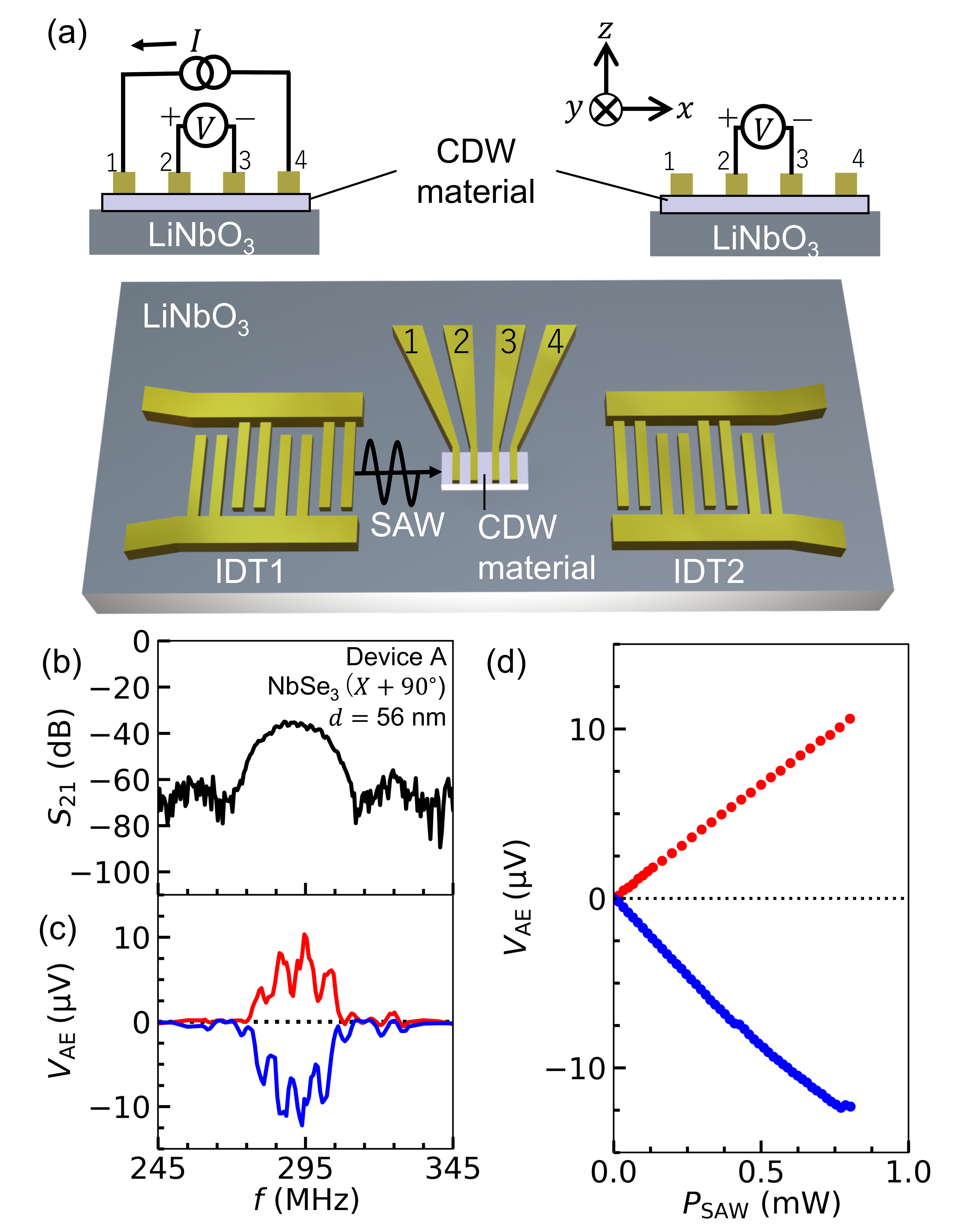}
	\end{minipage}
	\caption{
            (a) Schematic illustration of the SAW device. The top left and right figures are, respectively, for the resistance measurement and for the AE current measurement.
            (b),~(c) The scattering parameter $|S_{21}|$ (b) and the AE voltage $V_{\rm {AE}}$ (c) as a function of rf frequency. The AE voltage
            was measured for the SAW power ($P_{\rm SAW}$) of 0.82~mW. (d) $P_{\rm SAW}$ dependence of the AE current for $f=294$~MHz. All the results were obtained for NbSe$_3$($X+90^\circ)$ at $T=50$~K. In (c) and (d), the data for the rf signal supplied to IDT1 and IDT2 are shown in red and blue, respectively.
		\label{fig:setup}
	}
\end{figure}

NbSe$_3$ and 2H-TaSe$_2$ were selected as model systems to investigate the AE current in CDW states. NbSe$_3$ is a prototypical quasi-one-dimensional CDW compound that undergoes two CDW phase transitions at $\approx 145$~K and $\approx 60$~K~\cite{monceau1976}. The wave vector associated with the first CDW is oriented along the crystallographic $b$ axis, whereas that of the second CDW is inclined diagonally, extending across both the $a$–$c$ plane and the $b$ axis~\cite{schafer2001}. In reciprocal lattice units, the former and the latter are written as $(0,0.24,0)$ and $(0.5,0.26,0.5)$, respectively.
Thin NbSe$_3$ films preferentially align their long axis along the quasi-one-dimensional $b$ axis~\cite{fujiwara2021}. 
2H-TaSe$_2$ exhibits two-dimensional CDW transitions, characterized by an incommensurate CDW transition at $\approx 120$ K, followed by a commensurate CDW phase at $\approx 90$ K~\cite{barmatz1975}, whose wave vectors are $(0.32,0,0)$ and $(1/3,0,0)$, respectively~\cite{shen2023}.

Thin films of NbSe$_3$ and 2H-TaSe$_2$ were fabricated on a $128^\circ$ Y-cut black LiNbO$_3$ substrate via mechanical exfoliation using Scotch tape. For NbSe$_3$, flakes were selected such that their longer edges were closely aligned with the SAW propagation direction. In the case of 2H-TaSe$_2$, the crystallographic orientations of the exfoliated flakes were not determined in the present study. Device structures are schematically illustrated in Fig.~\ref{fig:setup}(a). Double-finger IDTs were employed to facilitate efficient SAW generation at the transmitting transducer and to minimize SAW reflection at the receiving transducer~\cite{bristol1972}. The IDTs and four-terminal electrical contacts were fabricated using electron-beam lithography, followed by the deposition of Ti(40~nm)/Au(30~nm) and a conventional lift-off process. Each IDT comprises 13 finger pairs with a pitch of $12.7\ \upmu$m, corresponding to the fundamental SAW wavelength $\lambda$. Throughout this study, a right-handed Cartesian coordinate system ($xyz$) is adopted, where the $x$ axis is defined parallel to the SAW propagation direction and the $z$ axis is normal to the substrate surface.

The film dimensions (thickness $d$, distance between voltage measurement electrodes $L$, and width $W$) employed in the AE current measurements are summarized in Table~\ref{table:devices}. Three types of devices were investigated, distinguished by the combination of SAW propagation direction and CDW material: $X+90^\circ$ with NbSe$_3$, $X+90^\circ$ with 2H-TaSe$_2$, and $X$ with NbSe$_3$. These devices are hereafter denoted as NbSe$_3$($X+90^{\circ}$), 2H-TaSe$_2$($X+90^{\circ}$), and NbSe$_3$($X$), respectively. Here, the $X+90^\circ$ direction is defined as the axis obtained by rotating the $X$ axis of LiNbO$_3$ by $90^\circ$~\cite{weser2020,kiefer2025}: see also Fig.~\ref{fig:iae0deg} (a). As for NbSe$_3$, three additional devices were investigated to confirm the reproducibility. See Appendix~\ref{sec:reproduce} for the results of Devices E, F, and G. In addition, as a control experiment, the AE current was also measured in a metallic Ta thin film deposited by rf magnetron sputtering. Two Ta devices with different length and width were prepared to investigate the possible size effect.
We believe that the flake thickness would be the most important length scale because it is relevant with the electrical transport properties including the temperature dependence of the resistivity and the threshold field of the CDW sliding~\cite{fujiwara2021}, which may affect the acoustoelectric properties.

The performance of the SAW devices was characterized through $S$-parameter measurements using a vector network analyzer (VNA, Keysight P9374B). A representative transmission coefficient ($S_{21}$) spectrum is presented in Fig.~\ref{fig:setup}(b), exhibiting a pronounced peak corresponding to efficient SAW excitation. A Gaussian fit to the data enables us to extract the center frequency $f_c$, from which the SAW velocity $v$ can be determined using the relation $v = f_c \lambda$. 
The evaluated SAW velocities are in good agreement with previously reported values for Rayleigh-type SAWs~\cite{weser2020,kawada2025jap}. Furthermore, we found that the peak amplitude of $S_{21}$ was little dependent on temperature ($T$), whereas the SAW velocity shows a slight increase at lower temperatures; see Appendix~\ref{sec:tdeps21} for further details.

\begin{table}[t]
    \caption{
        Summary of SAW device information.
    }
    \label{table:devices}
    {\renewcommand{\arraystretch}{1.5}
        \begin{tabular}{Wc{3em} Wc{4em} Wc{6em} Wc{3em} Wc{3em} Wc{4em} }
            \hline \hline
            Device & Material & Propagation & $d$ (nm) & $L$ ($\upmu$m) & $W$ ($\upmu$m) \\ \hline
            A & NbSe$_3$ & $X+90^\circ$ & 56 & 4.6 & 0.79 \\
            B & 2H-TaSe$_2$ & $X+90^\circ$ & 10 & 2.5 & 4.1 \\	
            C & Ta & $X$ & 14 & 39 & 100 \\
            D & NbSe$_3$ & $X$ & 25 & 3.0 & 0.69 \\
            E & NbSe$_3$ & $X+90^\circ$ & 69 & 3.1 & 0.96 \\
            F & NbSe$_3$ & $X$ & 25 & 2.9 & 0.77 \\
            G & NbSe$_3$ & $X$ & 49 & 3.8 & 0.98 \\
            H & Ta & $X$ & 14 & 3.0 & 0.88 \\
    \hline \hline
    \end{tabular}}
\end{table}

To investigate the acoustoelectric properties, temperature $T$ dependence of AE voltage $V_{\rm{AE}}$ was investigated~\cite{rotter1998}. The dc voltage generated in the flakes was measured as a function of the rf input frequency $f$ at each $T$. The sweep range of $f$ was chosen to be sufficiently broad to include the SAW's on-resonant frequency at all temperatures and was fixed during the $T$ dependence measurement. The flake edges were electrically open. We also performed the measurement in the closed circuit condition, which was consistent with the open circuit condition: see Appendix~\ref{sec:closed} for details. The Gaussian fit to the voltage spectrum determines the AE voltage amplitude denoted by $\tilde{V}_{\rm{AE}}$. The AE current $\tilde{I}_{\rm AE}$ was subsequently obtained by dividing $\tilde{V}_{\rm{AE}}$ by $-R$, where $R$ is the four-terminal resistance of the film, to eliminate the explicit dependence of $\tilde{V}_{\rm{AE}}$ on $R$: see Appendix~\ref{sec:closed} for details. The negative sign originates from the fact that the AE voltage represents a compensating voltage to the AE current. 

A lock-in detection technique was employed to enhance the signal-to-noise ratio~\cite{castilla2021}. Because the AE voltage is proportional to the SAW power, an rf signal with amplitude modulation (AM) was applied to the IDT, and the voltage component synchronized with the AM reference was detected using a lock-in amplifier. The AM depth and modulation frequency were set to 100\% and 173~Hz, respectively. Note that the $Y$ component of the lock-in signal was at least two orders of magnitude smaller than the $X$ component, indicating that the AM modulation frequency had little impact on the measured results.

The transport properties of the exfoliated flakes were evaluated from the $T$ dependence of the resistivity ($\rho$). To precisely evaluate the film resistance under SAW excitation, the four-terminal resistance was measured while applying an rf signal to the IDT. The signal frequency was set to the center frequency of the SAW peak, and the rf power and AM conditions were kept identical to those used for the AE voltage measurement. The resistance measurement was performed using a standard lock-in technique with an electric current of 1~$\upmu$A at a frequency of 37 Hz. This frequency was chosen to be sufficiently different from the AM frequency of the rf signal (173 Hz) to exclude the AE voltage. We remark that the film resistance change induced by SAW propagation was negligibly small; see Appendix~\ref{sec:wosaw} for details.


\section{Results}
\subsection{Fundamental acoustoelectric properties}
A representative AE voltage spectrum is shown in Fig.~\ref{fig:setup}(c), obtained for NbSe$_3$($X+90^{\circ}$) at 50 K. The signal is clearly observed within the SAW transmission band and exhibits a reversal in polarity upon inversion of the SAW propagation direction. Furthermore, as shown in Fig.~\ref{fig:setup}(d), the magnitude of the AE voltage increases linearly with the SAW power $P_{\rm SAW}$. Here, $P_{\rm SAW}$ indicates the actual SAW power delivered to the samples, which is estimated from the input rf power $P_{\rm in}$ and the transmission coefficient $|S_{21}|$. According to Ref.~\cite{fandan2019}, $P_{\rm SAW}$ is calculated as follows:
\begin{equation}
    P_{\rm SAW} = 10^{|S_{21}|\mathrm{[dB]}/20} P_{\rm in}.
    \label{eq:psaw}
\end{equation}
These observations are consistent with the established characteristics of the AE currents~\cite{wang1962,rotter1998,miseikis2012}. Note that the measurements in this work were performed within the range where the AE voltage is linearly proportional to $P_{\rm SAW}$. For the subsequent analysis, we focus on $\tilde{I}_{\rm AE}\equiv -V_{\rm{AE}}/R$ to normalize the explicit film resistance dependence of the AE voltage (see also Appendix~\ref{sec:closed}).

We remark that the oscillatory structure observed within the AE voltage peak [Fig.~\ref{fig:setup}(c)] is likely attributable to the oscillating AE effect~\cite{mou2025prl}. This phenomenon originates from the rectification of SAW-induced spatial charge modulation, driven by electromagnetic waves emitted from the IDTs. Because the oscillatory component is not essential to the analysis of the AE current, its contribution can be effectively removed by applying a Gaussian fit to the data. A more detailed discussion of the oscillating AE component is provided in Appendix~\ref{sec:oscae}.

\begin{figure}[tb]
	\begin{minipage}{1.0\hsize}
		\centering
		\includegraphics[scale=0.15]{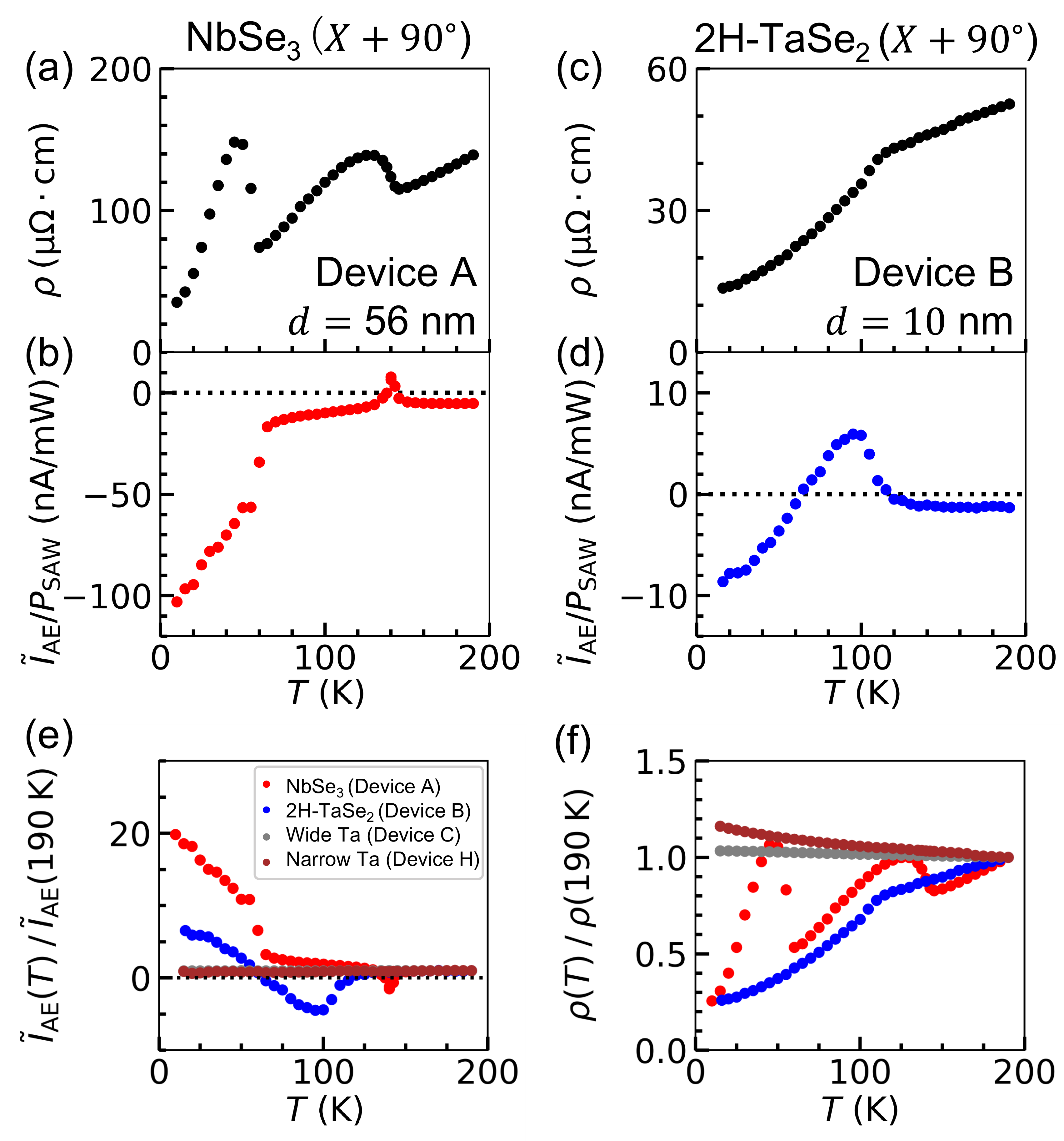}
	\end{minipage}
	\caption{(a),~(c)~Resistivity $\rho$ and (b),~(d)~$\tilde{I}_{\rm {AE}}$ obtained with NbSe$_3$($X+90^\circ)$ and 2H-TaSe$_2$($X+90^\circ)$ devices as a function of $T$. For (b) and (d), the SAW was irradiated from IDT1. (e),~(f)~$T$ dependence of (e)~$\tilde{I}_{\rm {AE}}$ and (f)~$\rho$ normalized by the values at $T=190$ K for NbSe$_3$ (red), 2H-TaSe$_2$ (blue), Ta (gray and brown) devices. 
		\label{fig:iae90deg}
	}
\end{figure}

\subsection{AE current in CDW states}
To investigate the acoustoelectric properties in the CDW states, the AE current was obtained as a function of $T$. Figures~\ref{fig:iae90deg}(a) and (b) show the $T$ dependence of the resistivity ($\rho$) and the AE current divided by the SAW power ($\tilde{I}_{\rm AE}/P_{\rm SAW}$) for NbSe$_3$($X+90^{\circ}$), respectively. To eliminate the SAW power dependence, the obtained AE current was normalized by the SAW power $P_{\rm SAW} = 0.82$~mW. As presented in Fig.~\ref{fig:iae90deg}(a), anomalies in $\rho$ are observed at $T_1 \approx 140$~K and $T_2 \approx 60$~K, corresponding to the two CDW transition temperatures~\cite{fujiwara2021}. Accordingly, the magnitude of the AE current exhibits a distinct peak at $T_{1}$ and a more pronounced increase below $T_{2}$. Notably, the polarity of the AE current around $T_1$ is positive, whereas it remains negative in other temperature regions.

An enhancement and polarity reversal of the AE current were also observed in 2H-TaSe$_2$($X+90^{\circ}$). The $T$ dependence of $\rho$, shown in Fig.~\ref{fig:iae90deg}(c), is consistent with previous reports~\cite{harper1977}. The first incommensurate CDW transition is identified as a weak kink at $T \approx 110$~K, whereas the subsequent commensurate CDW transition near 90~K is barely discernible in the $\rho$–$T$ curve~\cite{borisenko2008pseudogap}. Figure~\ref{fig:iae90deg}(d) presents the $T$ dependence of $\tilde{I}_{\rm AE}/P_{\rm SAW}$ obtained at $P_{\rm SAW} = 0.61$ mW. 
As the temperature decreases, the AE current begins to increase below $T\approx 120$~K, which would follow the first CDW transition. The AE current exhibits a single broad peak structure at $T\approx 95$~K and after that it monotonically decreases, which could be associated with the commensurate CDW transition. In contrast to the resistivity measurement that is sensitive to the density of states at the Fermi level, the AE current measurement may pick up the modulation of the conductivity by the strain.
The sign of $\tilde{I}_{\rm AE}$ is positive in the $T$ range from 115~K to 65~K, while it becomes negative outside this range.

Figure~\ref{fig:iae90deg}(e) compares the AE current for NbSe$_3$, 2H-TaSe$_2$, and Ta, normalized by their values at 190 K. In contrast to NbSe$_3$ and 2H-TaSe$_2$, the AE current in Ta is nearly independent of $T$ and does not exhibit any polarity reversal regardless of the film size. This observation indicates that the characteristic $T$ dependence and polarity change of the AE current are specific to CDW materials.
We remark that $T$ dependence of $\tilde{I}_{\rm AE}$ does not match that of $\rho$ as shown in Fig.~\ref{fig:iae90deg}(f).

\begin{figure}[tb]
	\begin{minipage}{1.0\hsize}
		\centering
		\includegraphics[scale=0.115]{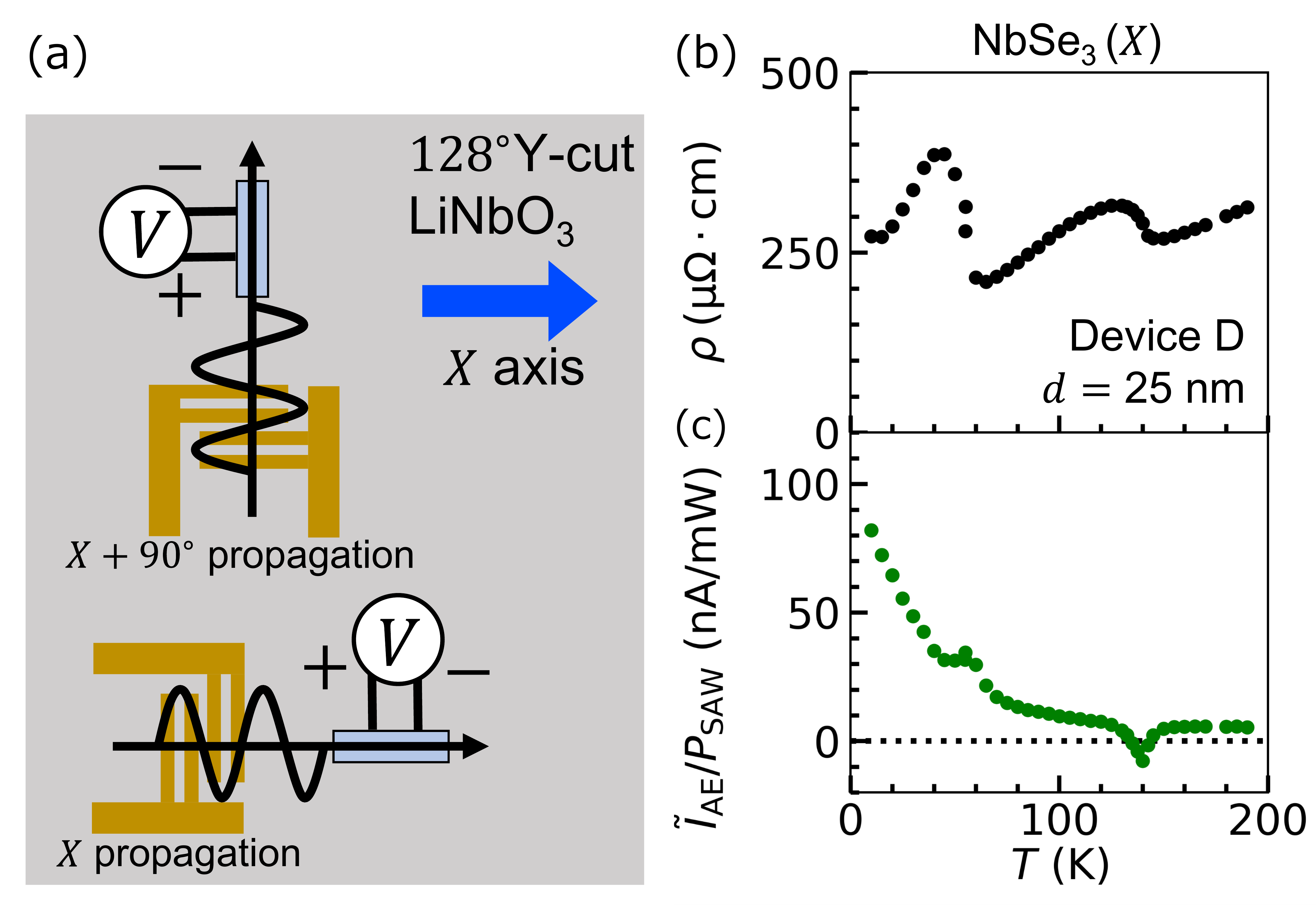}
	\end{minipage}
	\caption{(a) Schematic illustration of measurement geometries of $I_{\rm AE}$. The light gray square represents a $128^\circ$Y-cut LiNbO$_3$ substrate, where the blue arrow corresponds to the crystalline $X$ axis of LiNbO$_3$. The brown regions and the light blue rectangles indicate IDT1 and NbSe$_3$, respectively. (b),~(c) Temperature dependence of (b) $\rho$ and (c) $\tilde{I}_{\rm AE}$ for NbSe$_3$($X)$. The data in (c) were obtained with $P_{\rm SAW} = 0.1$~mW launched from IDT1.
		\label{fig:iae0deg}
	}
\end{figure}

\subsection{SAW propagation direction dependence}
A recent study demonstrated that the polarity of the AE current depends on the crystallographic direction of the substrate along which the SAW propagates~\cite{kawada2024exparxiv}. Motivated by Ref.~\cite{kawada2024exparxiv}, we fabricated an additional device, NbSe$_3$($X$), in which the SAW was launched along the $X$ axis of the $128^\circ$ Y-cut LiNbO$_3$ substrate. Importantly, the geometries of the voltage measurement terminals and the SAW propagation directions were kept identical between the two devices, as schematically illustrated in Fig.~\ref{fig:iae0deg}(a).

Figures~\ref{fig:iae0deg}(b) and \ref{fig:iae0deg}(c) show, respectively, $\rho$ and 
$\tilde{I}_{\rm AE}/P_{\rm SAW}$ (with $P_{\rm SAW}$ $=$ 0.10 mW) as a function of $T$. Similarly to the results for NbSe$_3$($X+90^{\circ}$), the AE current exhibits enhancements below the CDW transition temperatures and a sign change in the vicinity of the first CDW transition. However, the polarity of the AE current is opposite to that observed in NbSe$_3$($X+90^{\circ}$) over almost all the temperature range. As noted in the recent study~\cite{kawada2024exparxiv}, the conventional model~\cite{ingebrigtsen1970} fails to account for the polarity reversal of the AE current between the $X$ and $X+90^\circ$ propagation directions. The discrepancies between the conventional model and the present experimental results are further discussed in Appendix~\ref{sec:convmodel}.

\subsection{Model description}
The polarity reversal of the AE current can be understood by considering the rectification between the SAW-induced electric field and the strain field. As an illustrative example, we examine the rectification between the longitudinal strain $\varepsilon_{xx}$ and the transverse electric field $E_x$~\cite{kawada2024calcarXiv,kawada2024exparxiv,yamanaka2026}. The strain is expressed in a plane-wave form as $\varepsilon_{xx} = \varepsilon_{xx0} e^{i(kx - \omega t)}$, where $k$ and $\omega$ denote the wave number and angular frequency of the SAW, respectively. The corresponding electric field is given by $E_x = E_{x0} e^{i(kx - \omega t + \varphi)}$, where $\varphi$ represents the phase difference between $E_x$ and $\varepsilon_{xx}$. Under these definitions, the time-average of $\varepsilon_{xx} E_x$ is proportional to $\cos \varphi$. A previous study demonstrated that $\cos \varphi$ takes opposite signs for Rayleigh-type SAWs propagating along the $X$ and $X+90^\circ$ directions on a $128^\circ$ Y-cut LiNbO$_3$ substrate coated with a metallic thin film due to the anisotropic piezoelectric constants of LiNbO$_3$~\cite{kawada2024exparxiv}. Thus, the reversal of the AE current polarity between the $X$ and $X+90^\circ$ propagation directions can be consistently explained if the AE current originates from the rectification between $E_x$ and $\varepsilon_{xx}$.

Based on the above considerations, we propose a phenomenological model that can account for the polarity of the AE current. In the model, the longitudinal strain $\varepsilon_{xx}$ is assumed to perturbatively modulate the electrical conductivity of the CDW material, which can be expressed as
\begin{equation}
    \sigma (\varepsilon_{xx}) \approx \sigma_0 + \pdv{\sigma}{\varepsilon_{xx}}\varepsilon_{xx},
    \label{eq:sigmamod}
\end{equation}
where $\sigma_0$ represents the conductivity in the absence of strain. The strain-induced modulation of conductivity may arise, for example, from the piezoresistance effect~\cite{smith1954}. Concurrently, the electric field $E_x$ is applied to the flake, driving the conduction electrons. Note that the transverse electric field $E_x$ is screened not by the induced charges but by the electric current. It can almost uniformly penetrate into the flake if the films thickness is sufficiently smaller than the skin depth ($\sim10\: \upmu$m). Assuming a relation $J_x = \sigma E_x$, the time-averaged current density, integrated over the $y$ and $z$ directions yields the following expression for the AE current:
\begin{equation}
    \tilde{I}_{\rm AE} = \frac{Wd}{2}\cdot\frac{1}{\sigma_0}\pdv{\sigma}{\varepsilon_{xx}}J_0 \cos{\varphi},
    \label{eq:JAE}
\end{equation}
where $J_0$ is defined as $\sigma_0 \varepsilon_{xx0}E_{x0}$. While we assumed the electric current generation from the SAW here, the equivalent model can also be given by assuming the voltage generation: see Eqs.~(\ref{eq:rhomod}) and (\ref{eq:VAE}) in Appendix~\ref{sec:closed} for the detail.

The proposed model can qualitatively reproduce the experimental observations at least for NbSe$_3$. Previous studies have investigated the $T$ dependence of the resistivity in bulk NbSe$_3$ under static uniaxial strain, demonstrating that strain modifies the resistivity through shifts in the CDW transition temperatures~\cite{lear1984,tseng1995}. 
If the strain is comparable in magnitude to that induced by SAW, tensile strain would reduce the resistivity of NbSe$_3$ in the vicinity of $T_1$, while increasing it at other $T$ regions. This behavior implies that $\partial \sigma / \partial \varepsilon_{xx} > 0$ near $T_1$ and $\partial \sigma / \partial \varepsilon_{xx} < 0$ elsewhere, which can reproduce the observed sign reversal of the AE current around $T_1$. Furthermore, $(1/\sigma_0)(\partial \sigma / \partial \varepsilon_{xx})$ has been shown to enhanced below the CDW transition temperatures, particularly below $T_2$~\cite{lear1984,tseng1995}, suggesting the increase of the AE current after the CDW transitions. 

To the best of our knowledge, the strain dependence of the resistivity in 2H-TaSe$_2$ has not yet been reported. Nevertheless, it is plausible that the observed $T$ dependence of the AE current in 2H-TaSe$_2$ reflects a similar strain-induced modulation of its resistivity.

\section{Discussion}
We further examine the proposed scenario, i.e., Eq.~(\ref{eq:JAE}), through model calculations. $\varepsilon_{xx0}$, $E_{x0}$, and $\varphi$ were estimated by computing the SAW propagation characteristics based on established methods~\cite{ingebrigtsen1969,ingebrigtsen1970,datta1986,kushibiki1999ieee,kawada2025jap}. Details of the calculation are provided in Appendix~\ref{sec:calc}. As shown in Fig.~\ref{fig:iaetase2}(a), the quantity $(J_0/P_{\rm SAW})\cos \varphi$ exhibits a weak $T$ dependence. This indicates that the $T$ dependence of the AE current is primarily governed by that of the strain modulation of the conductivity. 

Assuming the validity of Eq.~(\ref{eq:JAE}), we evaluated $(1/\sigma_0)(\partial \sigma / \partial \varepsilon_{xx})$ from the AE current data in combination with the numerically obtained values of $J_0 \cos \varphi$. The results are presented in Figs.~\ref{fig:iaetase2}(b)–\ref{fig:iaetase2}(d). The absolute magnitude of $(1/\sigma_0)(\partial \sigma / \partial \varepsilon_{xx})$ reaches approximately $10^2$ for 2H-TaSe$_2$ and $10^3$–$10^4$ for NbSe$_3$. Note that these values are orders of magnitude larger than that for Ta: for Device C, we obtained $(1/\sigma_0)(\partial \sigma / \partial \varepsilon_{xx})\approx -2$.
The magnitude of $(1/\sigma_0)(\partial \sigma / \partial \varepsilon_{xx})$ for NbSe$_3$($X+90^\circ$) is larger than that for NbSe$_3$($X+90^{\circ}$) by a factor about four. 
The discrepancy between NbSe$_3$($X+90^\circ$) and NbSe$_3$($X$) may originate from sample variations in NbSe$_3$. Indeed, the transport properties differ among each flake, as exemplified by the quantitative differences observed in the $\rho$–$T$ characteristics (see Fig.~\ref{fig:iae90deg}(a) and Fig.~\ref{fig:iae0deg}(b)). Consequently, $(1/\sigma_0)(\partial \sigma / \partial \varepsilon_{xx})$ might also vary among devices.

While the enhancement and the polarity reversal of the AE current associated with the CDW transitions were common for both NbSe$_3$ and 2H-TaSe$_2$, the two CDW materials showed some quantitatively different AE properties. For instance, the peak width of $\tilde{I}_{\rm AE}$ for 2H-TaSe$_2$ was broader than that for NbSe$_3$. In addition, for 2H-TaSe$_2$, the ratio of the peak height (near 95~K) to the AE current at the lowest temperature was relatively larger than those (near 140~K) for NbSe$_3$. These differences might root from the detailed properties of specific CDW states such as their different band structures and dimensionalities (one- or two-dimensional CDWs). Studies on the strain response of the resistivity would shed light on the detail of the AE current in 2H-TaSe$_2$. 

Here we focus on NbSe$_3$, for which the piezoresistance has been experimentally investigated. The magnitude of $(1/\sigma_0)(\partial \sigma / \partial \varepsilon_{xx})$ obtained in the present study is substantially larger than the previously reported values, which were obtained against the static strain~\cite{lear1984,tseng1995}.
One possible origin of this discrepancy is the difference between thin-film and bulk forms of NbSe$_3$. It is known that NbSe$_3$ thin films exhibit transport properties distinct from those of bulk crystals~\cite{fujiwara2021}, suggesting that the strain-induced modulation of conductivity may also differ. Another factor is whether the applied strain is dc or ac. While previous studies considered static strain, the present work involves dynamic strain induced by SAWs. Previous studies on a related CDW material, TaS$_3$, have shown that CDW transport can be more strongly modulated by dynamic strain than by static strain~\cite{nikitin2021}. This observation implies that $(1/\sigma_0)(\partial \sigma / \partial \varepsilon_{xx})$ may attain significantly larger values under ac strain compared to dc strain.

Within the present experimental condition, the measurements were performed in the linear regime where the CDW dynamics does not occur, which is implied by the following results. First, the AE current was linearly proportional to $P_{\rm SAW}$, indicating that the resistance remained in a linear response regime. Second, the electric field induced by the AE voltage was far below the threshold field in thin films: for instance, the former is at most $0.01$~V/cm, whereas the latter reaches $10$~V/cm for NbSe$_3$ thin films~\cite{fujiwara2021,fujiwara2025prl}. Third, the $T$ dependence of the resistivity with and without SAW irradiation shows negligible differences (see Fig.~\ref{fig:rhoT} in Appendix~\ref{sec:wosaw}). Fourth, the ac resistance was in good agreement with its dc resistance (see Fig.~\ref{fig:acmeas} in Appendix~\ref{sec:calc} for the result of NbSe$_3$). The last two indicate that the CDW was not depinned by the SAW. These results suggest that the CDW electrons was not driven by the SAW in this work.

\begin{figure}[tb]
	\begin{minipage}{1.0\hsize}
		\centering
		\includegraphics[scale=0.16]{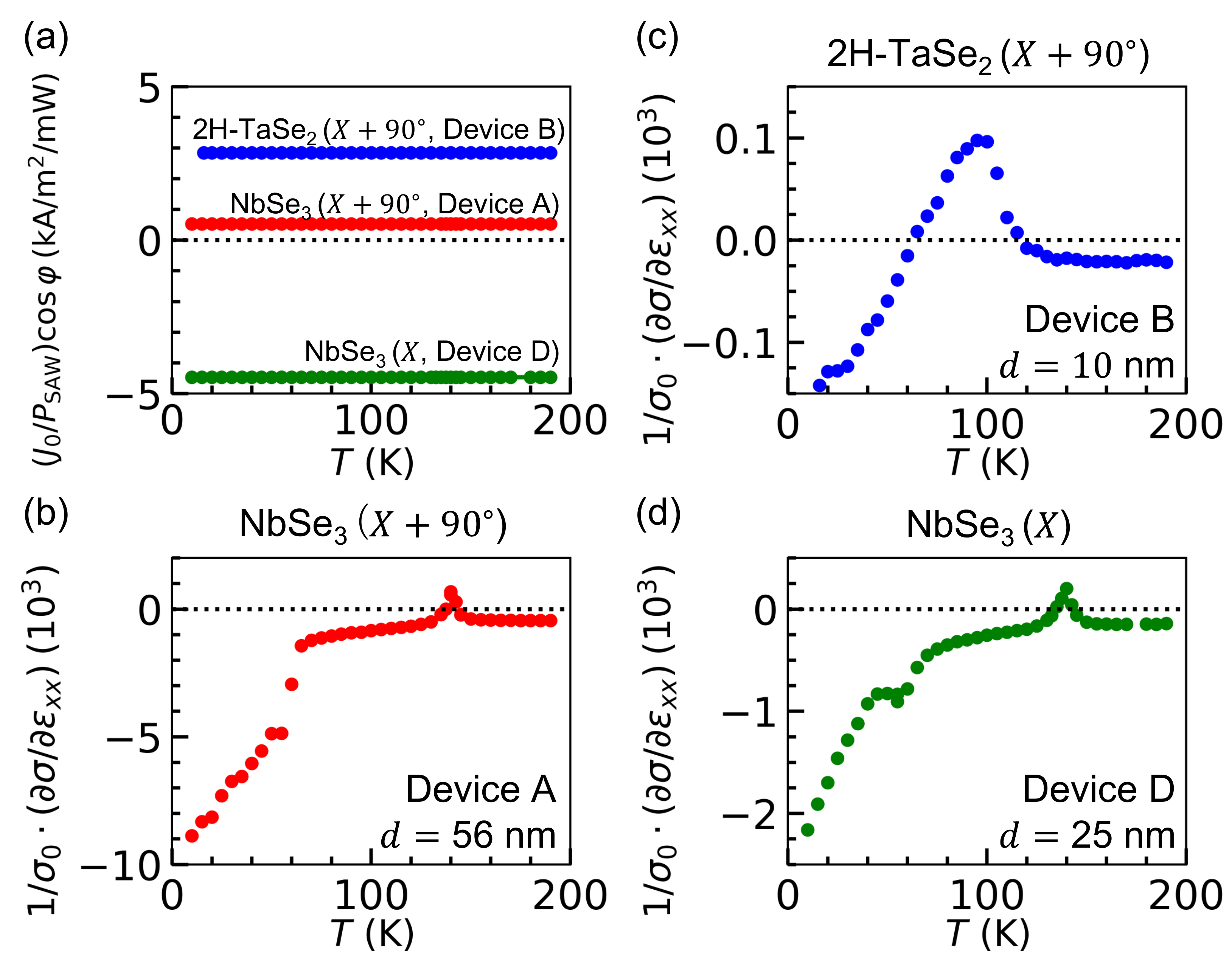}
	\end{minipage}
	\caption{
    (a) Numerically calculated result of $(J_0/P_{\rm SAW})\cos \varphi$. The results for NbSe$_3$($X+90^{\circ}$), 2H-TaSe$_2$($X+90^{\circ}$), and NbSe$_3$($X$) are shown in red, blue, and green, respectively. (b)-(d) Estimation of $(1/\sigma_0)(\partial \sigma/ \partial \varepsilon_{xx})$ for NbSe$_3$($X+90^{\circ}$) (b), 2H-TaSe$_2$($X+90^{\circ}$) (c), and NbSe$_3$($X$) (d), respectively.
	\label{fig:iaetase2}
	}
\end{figure}


The application of an rf signal to the IDTs may generate a thermal gradient along the $x$ direction, potentially leading to a thermoelectric voltage via the Seebeck effect. However, this contribution is expected to be negligible in the present case, as the dc voltage measured at off-resonant frequencies is significantly smaller than that observed at the resonant frequency.

The local heating due to the AE current passing in CDW flakes can be negligibly small. The Joule heating due to the AE current is at most $10^{-15}$ J in the present work. Assuming a typical sample volume of NbSe$_3$ ( $10\:\rm{\upmu m} \times 1\:\rm{\upmu m} \times 50\:\rm{nm}$), the mass density of 6.4 g/cm$^3$~\cite{kikkawa1982}, and the heat capacity of 3 mJ/g/K~\cite{lasjaunias1982}, the Joule heat can increase the sample temperature by $\sim 0.1$ K, which is at least two orders of magnitude smaller than the sample temperature.

While several theoretical models have been proposed to explain dc electric current generations from acoustic waves so far~\cite{ingebrigtsen1970,kalameitsev2019,sukhachov2020,ogata2026}, not all the present experimental results, including the temperature dependence and propagation-direction dependence of the AE current in the CDW materials, have been consistently explained by those models. This remains an issue for future work.

\section{Conclusion}
In summary, we have observed a polarity reversal and enhancement of the acoustoelectric (AE) current associated with charge-density-wave (CDW) transitions in NbSe$_3$ and 2H-TaSe$_2$, which are exfoliated on a piezoelectric LiNbO$_3$ substrate. A $90^\circ$ rotation of the surface acoustic wave (SAW) propagation direction results in a reversal of the AE current polarity, indicating that both the electric field and strain components of the SAW contribute to the generation of the AE current. We have proposed a phenomenological model in which the AE current arises from the rectification between the SAW-induced electric field and strain. This model accounts for the AE current polarity and its enhancement associated with the CDW transitions in NbSe$_3$, assuming that the conductivity is modulated by strain. Notably, the magnitude of the strain-induced conductivity modulation is several orders of magnitude larger than that reported in previous studies. This discrepancy may originate from the thin-film nature of the CDW materials and/or the influence of dynamic strain. These findings provide a powerful approach for probing SAW–electron interactions in van der Waals materials and may contribute to the advancement of straintronics.

\begin{acknowledgments}
We thank K.~Aoyama, Y.~Funami, K.~Yamamoto, S.~Maekawa, M.~Mori, H.~Matsukawa, M.~Hayashi, and M.~Ogata for the fruitful discussions. This work was supported by JSPS KAKENHI (Grant Nos. JP22KJ2180, JP22J20076, JP23H00257, JP23KJ1419, JP24K23028, JP25H01613, JP25H02122, JP25K17907), JST FOREST (Grant No.~JPMJFR2134), and the Cooperative Research Project of RIEC, Tohoku University.
\end{acknowledgments}

\appendix
\section{Reproducibility\label{sec:reproduce}}
\begin{figure}[t]
	\begin{minipage}{1.0\hsize}
		\centering
		\includegraphics[scale=0.155]{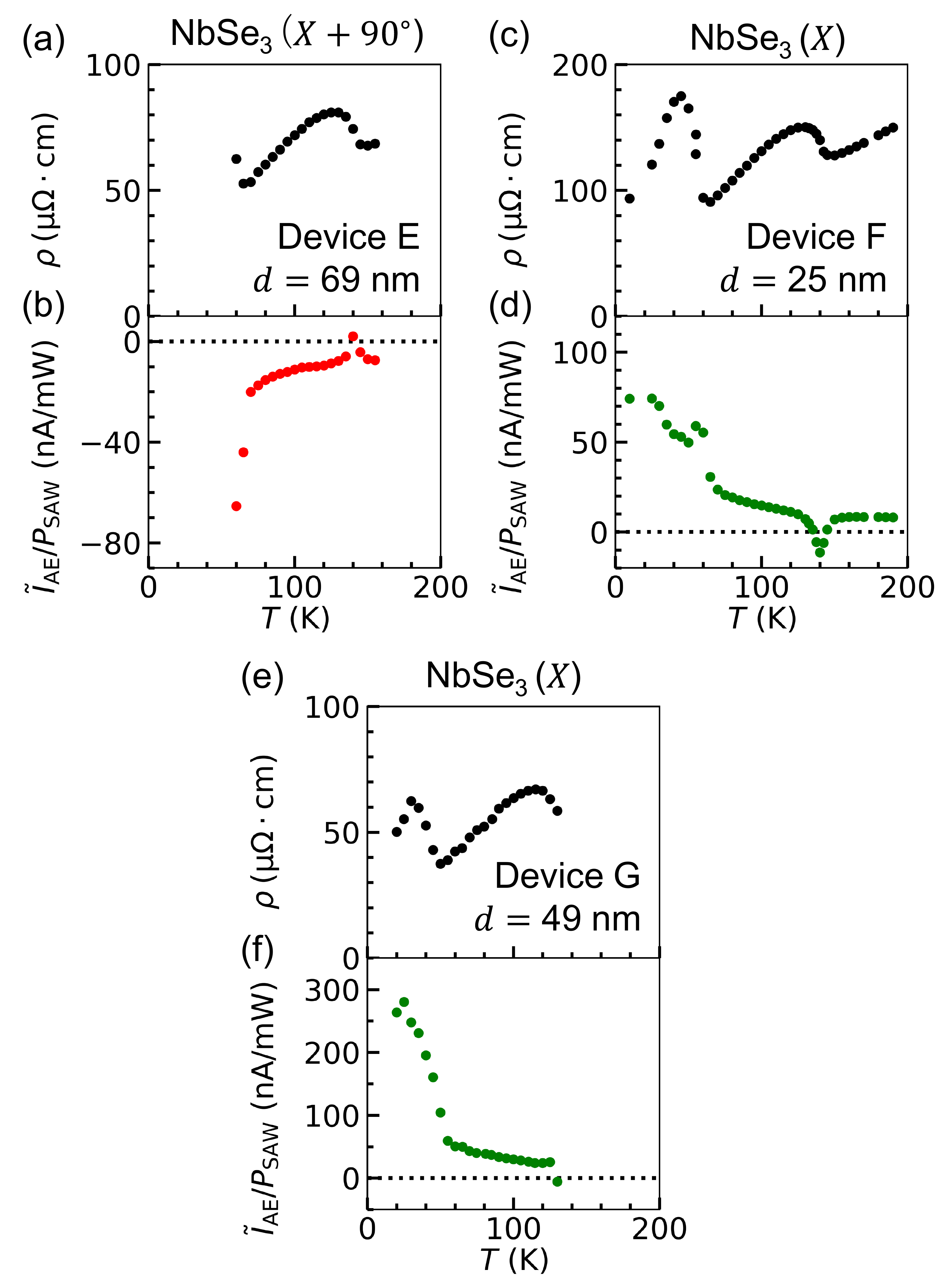}
	\end{minipage}
	\caption{
    (a),~(c),~(e)~Resistivity $\rho$ and (b),~(d),~(f)~$\tilde{I}_{\rm {AE}}$ obtained for (a),~(b)~Device E, (c),~(d)~Device F, and (e),~(f)~Device G as a function of $T$. For (b),~(d),~(f), the SAW was irradiated from IDT1.  
    \label{fig:rep}
	}
\end{figure}
Modulation of the AE current in the CDW state was confirmed in several SAW devices with NbSe$_3$. 
Figure~\ref{fig:rep} shows the $T$ dependence of the resistivity and the AE current for Device E, F, and G. The SAW propagation direction is $X+90^{\circ}$ for Device E, and $X$ for Device F and G. Note that Devices E and G were broken by the electrostatic discharging during the measurements, which hindered obtaining data at the other temperatures. Their AE current data exhibit quantitatively the same behavior as those of Device A and D. The AE current magnitude slightly and largely increases after the first and second CDW transitions. The polarity of the AE current around $T_1$ is opposite to that in the other temperature region. The AE current for $X$-propagation takes the opposite sign to that for $X+90^{\circ}$-propagation. Comparing the results of Devices D, F, and G implies that the NbSe$_3$ thickness does not affect the polarity of the AE current. These results further support the AE current mechanism proposed in this work.

\section{Temperature dependence of $S_{21}$\label{sec:tdeps21}}
\begin{figure}[bht]
	\begin{minipage}{1.0\hsize}
		\centering
		\includegraphics[scale=0.16]{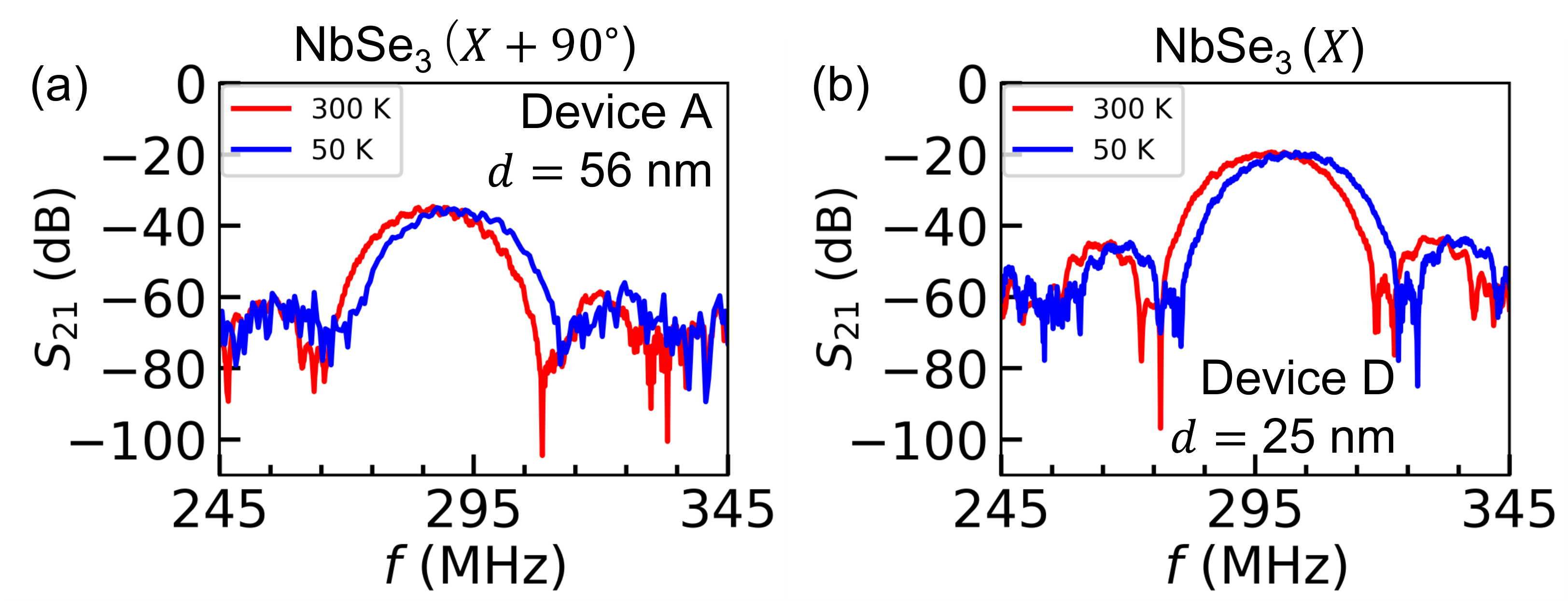}
	\end{minipage}
	\caption{Rf frequency dependence of $S_{21}$ (a)~for NbSe$_3$($X+90^\circ)$ and (b)~for NbSe$_3$($X)$. Red and blue curves show the data for 300~K and 50~K, respectively.
    \label{fig:s21}
	}
\end{figure}

The $S_{21}$ spectra measured at 300~K and 50~K are presented in Fig.~\ref{fig:s21}. The amplitude of $S_{21}$ is found to be nearly temperature-independent, indicating that the SAW power $P_{\rm SAW}$ is little dependent on $T$. The SAW velocity slightly increases with decreasing temperature, consistent with previous reports~\cite{kim1974}, and can be attributed to the stiffening of the elastic constants of LiNbO$_3$ at low temperatures~\cite{tarumi2012low}. The $S_{21}$ characteristics are not affected by the CDW flakes, as the SAW propagation region (approximately $100\ \upmu$m in width and $1$ mm in length) is significantly larger than the area occupied by the flakes.

\section{Acoustoelectric measurement under closed circuit condition\label{sec:closed}}

\begin{figure}[b]
	\begin{minipage}{1.0\hsize}
		\centering
		\includegraphics[scale=0.09]{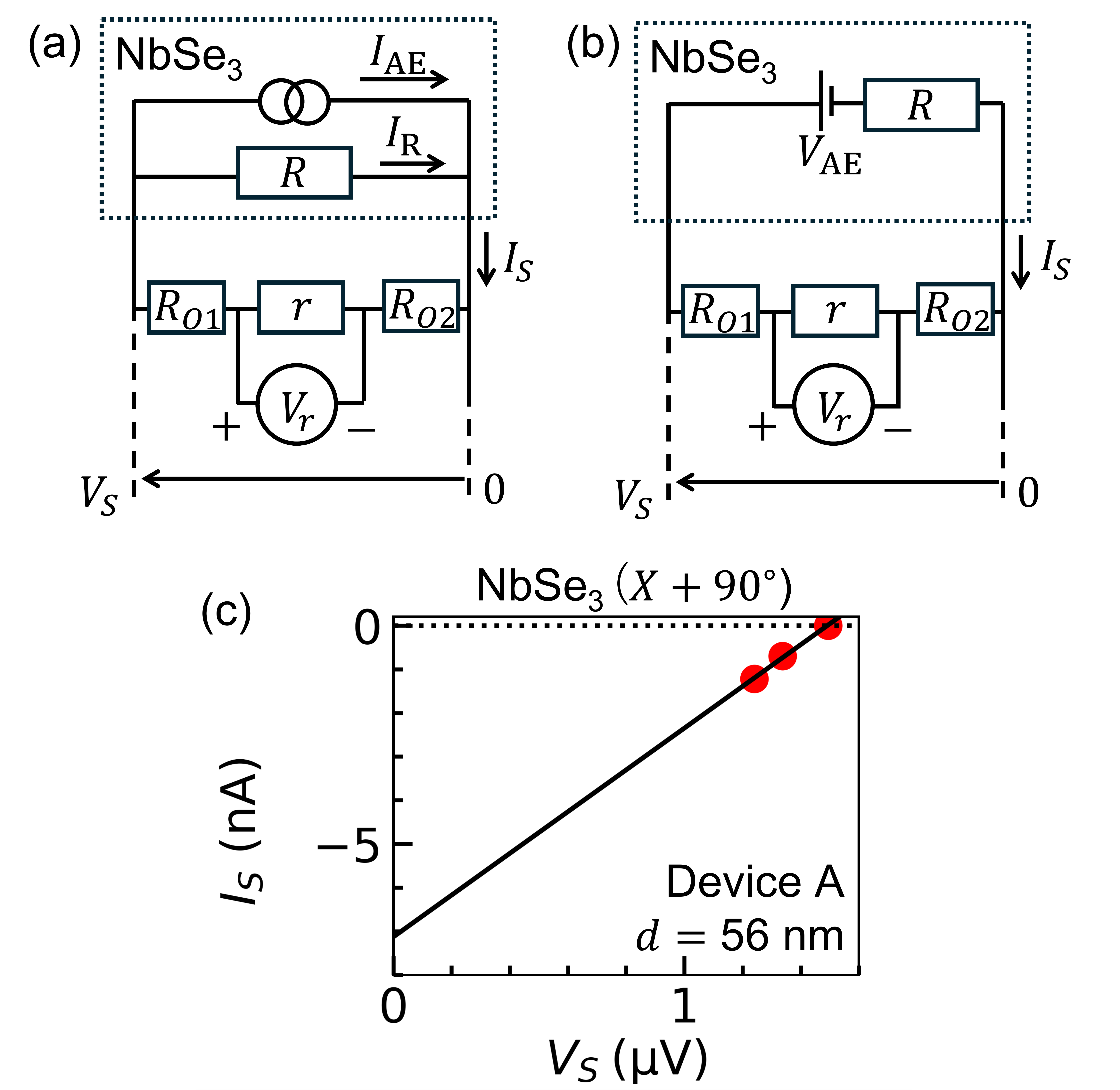}
	\end{minipage}
	\caption{(a),~(b)~Schematic illustrations of the closed circuit condition. The area inside the dotted square corresponds to NbSe$_3$. (c) The electric current outside NbSe$_3$ ($I_S$) plotted against the film voltage ($V_S$). The data were obtained for NbSe$_3$($X+90^\circ$) at $T=300$~K using the SAW of $P_{\rm SAW} =0.82$~mW launched from IDT1. The solid line shows a linear fit to the data. 
    \label{fig:ivmeas}
	}
\end{figure}
Due to acoustoelectric interactions, the SAW induces a dc electric current or voltage in the shorted or open circuits~\cite{rotter1998}.
In this study, we employed the open circuit condition because non-negligible resistances of equipment such as those of wires and electrical contacts prevented obtaining the shorted circuit condition with high accuracy; in the present case, the wire resistance was about 100~$\Omega$, and the contact resistance between NbSe$_3$ and Ti/Au electrode was about 300~$\Omega$, which are non-negligible compared with the film resistance.  

We also investigated the AE response using the closed circuit condition.
A reference resistor was connected to the ends of the flake, enabling the AE current to be partially extracted outside the sample. The relationship between the sample voltage and the electric current flowing in the circuit was then evaluated. The corresponding equivalent circuit is schematically illustrated in Fig.~\ref{fig:ivmeas}(a). The relevant quantities are defined as follows: the film resistance $R$, the reference resistance $r$, the sum of wire and contact resistances $R_{\rm O1}$ and $R_{\rm O2}$, the AE current $I_{\rm AE}$, the compensating current $I_R$, the current flowing outside the sample $I_S$, the voltage across the reference resistor $V_r$, and the sample voltage $V_S$. The sign of the currents is chosen such that $I_R$ is positive when $V_S$ is positive. 

Experimentally, $V_r$ was measured to determine $I_S$ for various values of $r$ using the relation $I_S = -V_r / r$. The corresponding sample voltage was then evaluated as $V_S = -I_S (R_{\rm O} + r)$. The resistance $R_{\rm O}\equiv R_{\rm O1} + R_{\rm O2}$ was obtained by subtracting the four-terminal resistance $R$ from the measured two-terminal resistance. The relationship of $I_S$ and $V_S$ can be modeled by the following formula:
\begin{equation}
    I_S = I_{\rm AE} + \frac{V_S}{R}.
    \label{eq:imodel_final}
\end{equation}
We note that the same relationship also holds for the circuit shown in Fig.~\ref{fig:ivmeas}(b), indicating that the AE current is equivalent to the combination of the AE voltage with an output resistance $R$ within the scope of this work. The separability of these two circuit models requires further study.

Figure~\ref{fig:ivmeas}(c) shows $I_S$ plotted as a function of $V_S$ for NbSe$_3$($X+90^\circ$) at $300$ K. $R_{\rm O}$ was determined to be $914~\Omega$, and $P_{\rm SAW}$ was set to 0.82 mW. We found that $I_S$ exhibits a linear dependence on $V_S$. Then we fit the data by a linear function:
\begin{equation}
    I_S = I_{\rm fit} + \frac{V_S}{R_{\rm fit}}.
    \label{eq:imodel_fit}
\end{equation}
The fitting yields $I_{\rm fit} = -7.1 \pm 0.3~\mathrm{nA}$ and $R_{\rm fit} = 210 \pm 10~\Omega$, where the deviation indicates the fitting error. These values are in good agreement with the AE current obtained from the voltage measurement ($I_{\rm AE} = -6.9~\mathrm{nA}$) and the sample resistance determined by the four-probe method ($R = 216~\Omega$). This agreement indicates that, within experimental accuracy, the open and closed circuit conditions provide quantitatively the same results. 

When the reference resistor is removed, the condition $I_S = 0$ is satisfied. Under this condition, Eq.~(\ref{eq:imodel_final}) reduces to $I_{\rm AE} = -V/R$, where $V$ denotes the AE voltage. This relation validates the evaluation of the AE current through dc voltage measurements.

Note that while we employed the electric current generation model in Eqs.~(\ref{eq:sigmamod}) and (\ref{eq:JAE}), a voltage generation model also leads to an equivalent model equation. Suppose that the longitudinal strain $\varepsilon_{xx}$ modulates the electrical resistivity of the CDW material as 
\begin{equation}
    \rho (\varepsilon_{xx}) \approx \rho_0 + \pdv{\rho}{\varepsilon_{xx}}\varepsilon_{xx},
    \label{eq:rhomod}
\end{equation}
where $\rho_0=\sigma_0^{-1}$ represents the resistivity in the absence of strain. The electric current $J_x$, which is generated from the electric field $E_x$ according to $J_x = \sigma E_x$, simultaneously flows in the flake. Assuming a relation $E_x = \rho J_x$, the time-averaged electric field integrated over the $x$ direction yields the following expression for the AE voltage:
\begin{equation}
    \tilde{V}_{\rm AE} = \frac{L\rho_0}{2}\cdot\frac{1}{\rho_0}\pdv{\rho}{\varepsilon_{xx}} J_0 \cos{\varphi},
    \label{eq:VAE}
\end{equation}
where $J_0$ is defined as $\sigma_0 \varepsilon_{xx0}E_{x0}$. Considering the relation of $(1/\rho_0)(\partial \rho / \partial \varepsilon_{xx})=-(1/\sigma_0)(\partial \sigma / \partial \varepsilon_{xx})$, $\tilde{V}_{\rm AE}$ and $\tilde{I}_{\rm AE}$ are equated by $\tilde{I}_{\rm AE} = - \tilde{V}_{\rm AE} / R$, which therefore reproduces the results in this work. We remark that since $J_0$ is independent of the film resistivity as shown in Fig.~\ref{fig:calc} (a) in Appendix~\ref{sec:calc}, the AE voltage is proportional to $\rho_0$ while the AE current is not explicitly
dependent on $\rho_0$.

\section{Temperature dependence of resistivity with and without SAWs\label{sec:wosaw}}

\begin{figure}[tbh]
	\begin{minipage}{1.0\hsize}
		\centering
		\includegraphics[scale=0.2]{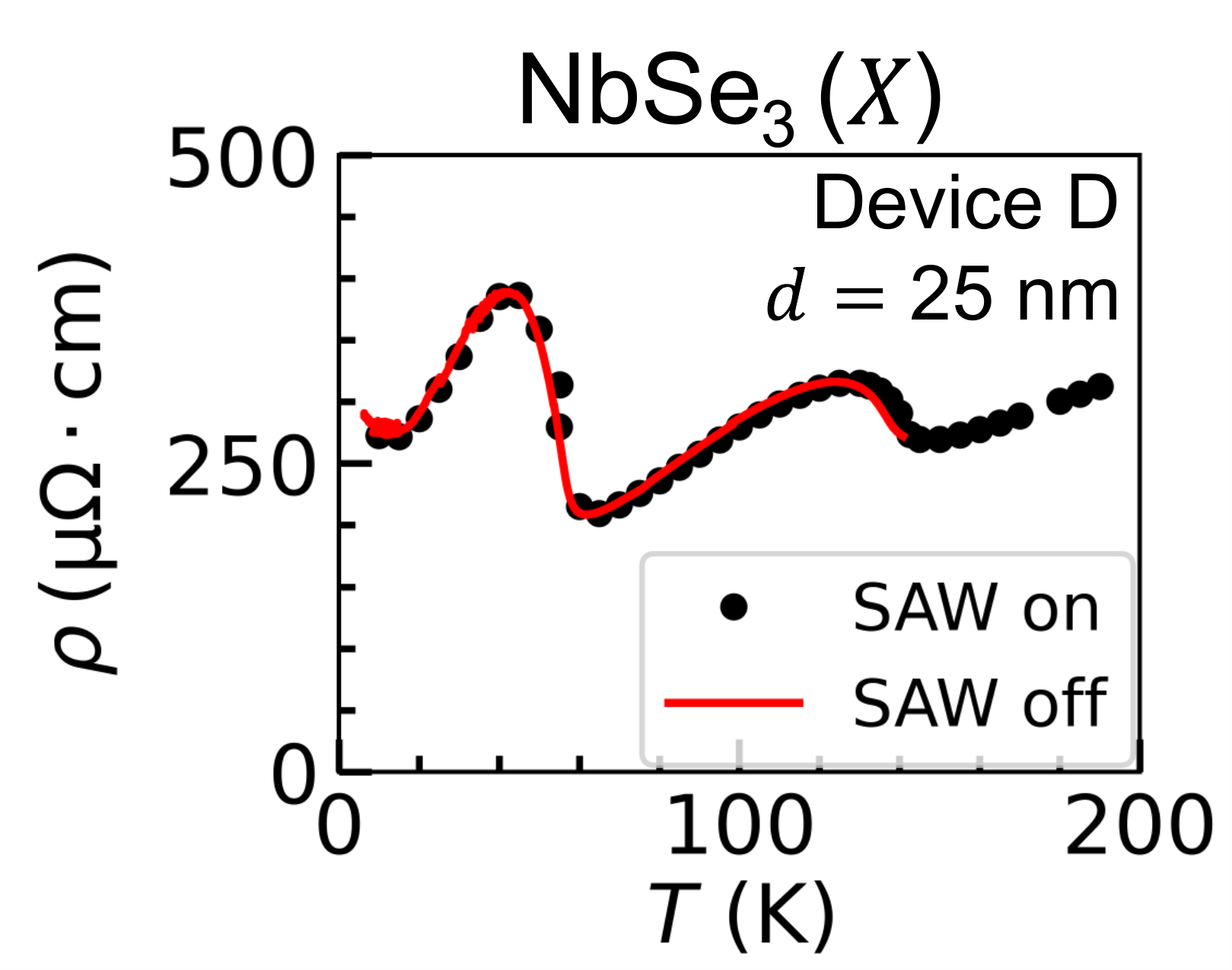}
	\end{minipage}
	\caption{ Temperature dependence of $\rho$ for NbSe$_3$($X)$. Black dots and red curves represent the data with and without SAW, respectively. Black dots are the same data as those shown in Fig.~\ref{fig:iae0deg}(b).
    \label{fig:rhoT}
	}
\end{figure}

In principle, the irradiation of the SAW may modulate the dc conductivity of NbSe$_3$ through sample heating and/or depinning of the CDWs. To examine these effects, we measured the $T$ dependence of the dc resistivity both in the presence and absence of SAW excitation. The results for NbSe$_3$($X$) are shown in Fig.~\ref{fig:rhoT}, obtained with and without SAW irradiation at $P_{\rm SAW} = 0.10$ mW. The horizontal shift of the $\rho$–$T$ curve is negligibly small, indicating that SAW-induced heating is insignificant. Furthermore, the two $\rho$–$T$ curves overlap within experimental uncertainty, suggesting that the CDWs are not depinned by the SAW under the present conditions.

\section{Oscillating components of the AE current\label{sec:oscae}}
The AE current spectra contained an additional oscillatory component, as exemplified in Fig.~\ref{fig:setup}(c) for NbSe$_3$($X+90^{\circ}$). This feature is likely attributable to the oscillating AE effect~\cite{mou2025prl}, which arises from the rectification of spatiotemporal carrier oscillations induced by the SAW via electromagnetic waves directly radiated from the IDTs.

To investigate the oscillating AE current, we fitted the following function to the AE current spectra:
\begin{equation}
    I_{\rm AE} = \Bigg[\tilde{I}_{\rm AE} + \tilde{I}_{\rm osc} \cos \Big(2\pi{\frac{f}{f_p}}+\phi_0 \Big)\Bigg] \ e^{-\frac{(f-f_c)^2}{2\Delta^2}}.
    \label{eq:psaw}
\end{equation}
In the above function, $\tilde{I}_{\rm AE}$ and $\tilde{I}_{\rm osc}$ denote the amplitudes of the AE current and the oscillatory AE component, respectively. The parameters $f_p$ and $\phi_0$ represent the oscillation period and the initial phase of the oscillatory component, while $f_c$ and $\Delta$ correspond to the center frequency and the spectral width of the AE current peak, respectively. Note that the values of $\tilde{I}_{\rm AE}$ obtained from this analysis are in good agreement with those extracted from a Gaussian fit to the experimental data.

The period $f_p$ is given by $f_p = v/l$, where $l$ is the distance between the centers of the IDT and the flake. We obtained $v = 3640$ m/s and $l = 456~\upmu\mathrm{m}$ for NbSe$_3$($X+90^{\circ}$), $v = 3580$ m/s and $l = 846~\upmu\mathrm{m}$ for 2H-TaSe$_2$($X+90^{\circ}$), and $v = 3780$ m/s and $l = 547~\upmu\mathrm{m}$ for NbSe$_3$($X$), yielding $f_p = 8.0$ MHz, 4.2 MHz, and 6.9 MHz, respectively.

\begin{figure}[t]
	\begin{minipage}{1.0\hsize}
		\centering
		\includegraphics[scale=0.15]{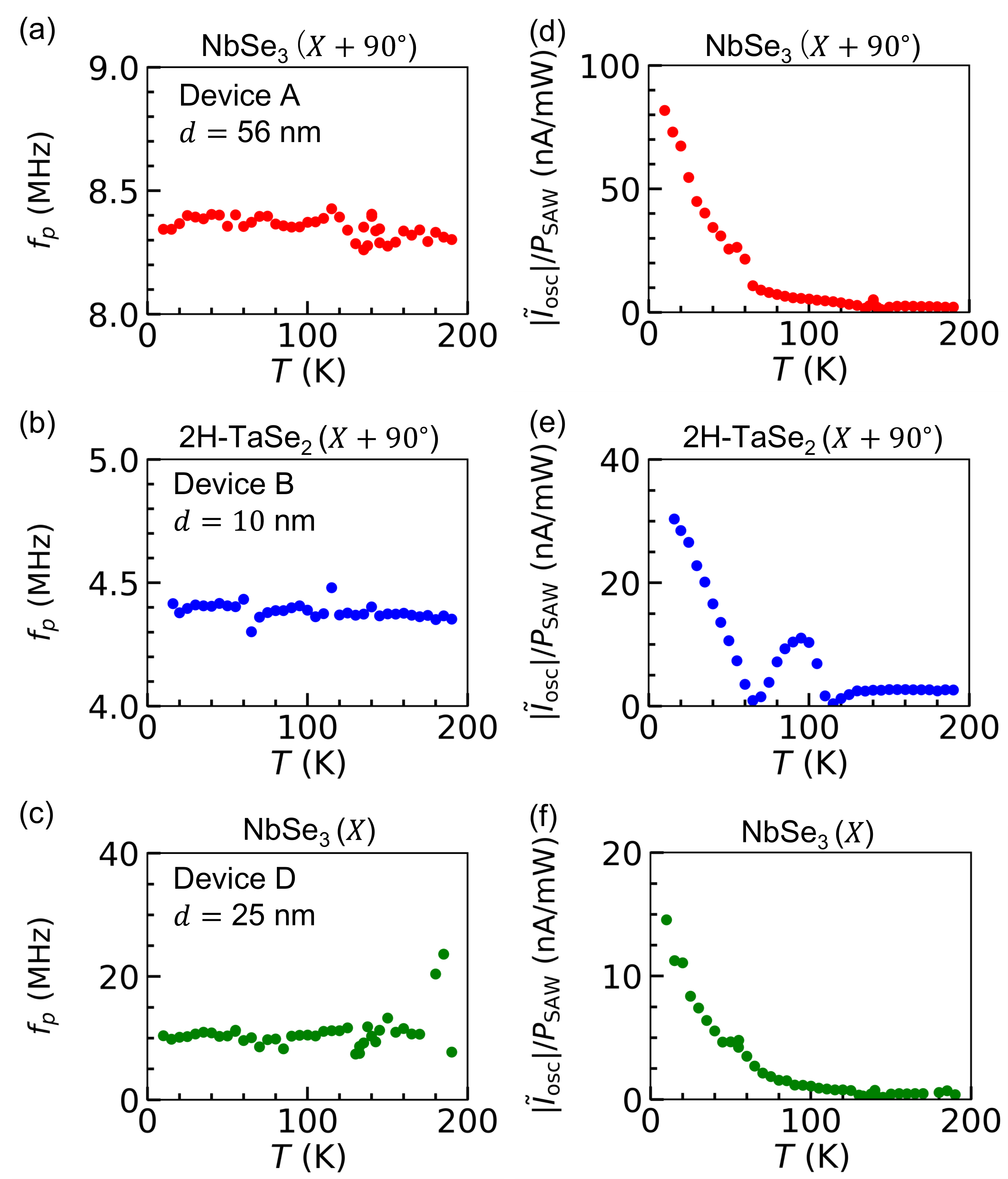}
	\end{minipage}
	\caption{(a),~(d) Temperature dependence of $f_p$ and $\tilde{I}_{\mathrm{osc}}/P_{\mathrm{SAW}}$ for NbSe$_3$($X+90^{\circ}$), (b),~(e) for 2H-TaSe$_2$($X+90^{\circ}$), and (c),~(f) for NbSe$_3$($X$). The SAW was propagated from the IDT1 to IDT2. \label{fig:osc}
	}
\end{figure}

Figures~\ref{fig:osc}(a)–(c) present the $T$ dependence of $f_p$. The values of $f_p$ for NbSe$_3$($X+90^{\circ}$) and 2H-TaSe$_2$($X+90^{\circ}$) are in good agreement with the model predictions, supporting the contribution of the oscillating AE effect. 
$f_p$ slightly increases towards low temperature, which is consistent with the increase of $v$ with decreasing $T$ (see also Fig.~\ref{fig:s21}). Small discrepancy between the predicted and observed $f_p$ may originate from slight deviations of $l$ and $\lambda$ from the designed values. Note that $f_p$ for NbSe$_3$($X$) differs from the model prediction, possibly because the relatively small magnitude of $\tilde{I}_{\mathrm{osc}}$ compared to $\tilde{I}_{\mathrm{AE}}$, which is less than one-fifth (see Fig.~\ref{fig:iae0deg}(c) and Fig.~\ref{fig:osc}(f)), hinders the precise determination of $f_p$.
The $T$ dependence of $|\tilde{I}_{\mathrm{osc}}|/P_{\mathrm{SAW}}$ is shown in Figs.~\ref{fig:osc}(d)–(f). Here, the sign of $\tilde{I}_{\mathrm{osc}}$ is not discussed, as the initial phase $\phi_0$, which also determines the sign, could not be accurately evaluated due to the limited frequency resolution in the measurements. It is found that $|\tilde{I}_{\mathrm{osc}}|/P_{\mathrm{SAW}}$ exhibits a similar $T$ dependence to that of $|\tilde{I}_{\mathrm{AE}}|/P_{\mathrm{SAW}}$, implying that the oscillatory AE current may share a common origin with the primary AE current in the present system, i.e., the SAW strain causes the spatiotemporal carrier oscillations.

\section{Conventional AE current model for NbSe$_3$\label{sec:convmodel}}

\begin{figure}[b]
	\begin{minipage}{1.0\hsize}
		\centering
		\includegraphics[scale=0.16]{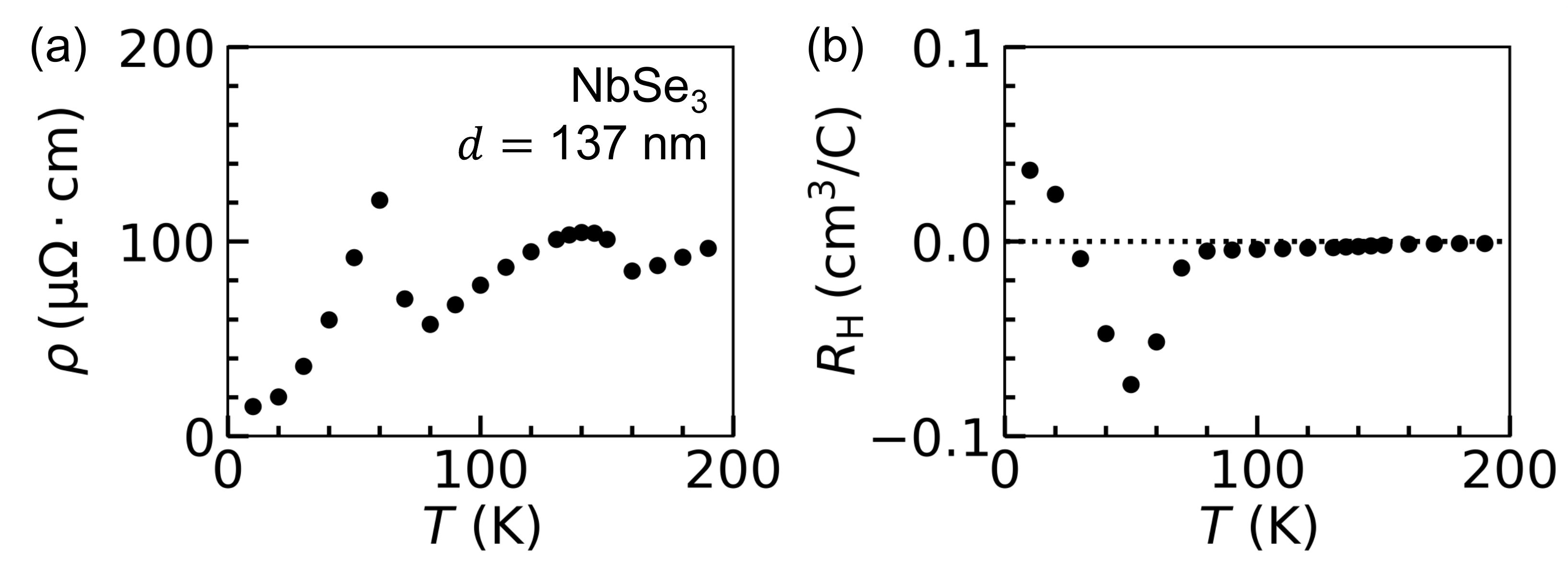}
	\end{minipage}
	\caption{Temperature dependence of (a) $\rho$, and (b) the low-field Hall coefficient $R_{\rm{H}}$ for 137 nm-thick NbSe$_3$ on a Si/SiO$_x$ substrate. $R_{\rm{H}}$ was obtained from the linear fitting to the data for the magnetic field ranging from $-1$ T to $1$ T.
	\label{fig:Hall}
	}
\end{figure}

The generation of AE current by piezoelectrically excited SAWs has been conventionally described by the model proposed by Ingebrigtsen~\cite{ingebrigtsen1970}. In this framework, the AE current is expressed as the time average of the product of the surface charge density and the carrier velocity. The latter is given by the product of the electric field and the carrier mobility. The original formulation was developed for conducting films with a single type of charge carrier. It was later extended to systems with two types of carriers possessing nearly identical mobilities, such as graphene~\cite{nichols2024}. However, these formulations cannot be directly applied to materials in which electrons and holes with different mobilities coexist. Indeed, NbSe$_3$ and 2H-TaSe$_2$ fall into this category of materials~\cite{ong1978,ong1978prb,naito1982}. 

Figure~\ref{fig:Hall}(b) shows the $T$ dependence of the Hall coefficient $R_{\mathrm{H}}$ for a NbSe$_3$ flake with a thickness of 137 nm, exfoliated onto a thermally oxidized silicon substrate. Over the entire temperature range investigated, the Hall resistance exhibited a linear dependence on the applied perpendicular magnetic field within $\pm 1$~T. Notably, $R_{\mathrm{H}}$ changes sign in the temperature range between 20 K and 30 K, indicating the coexistence of electron and hole carriers in NbSe$_3$ thin films. These observations confirm that transport in NbSe$_3$ is governed by a two-carrier mechanism, and therefore, the direct application of the conventional model equations is not appropriate for the present study.

Following the framework of the conventional model~\cite{ingebrigtsen1969,ingebrigtsen1970}, we derive an approximate expression for the AE current per unit length ($J_{\rm AE}$) in a metallic film containing two types of charge carriers as follows:
\begin{equation}
    J_{{\rm AE}} \approx \tilde{\mu} \frac{\tilde{\Gamma} I_{\rm SAW}}{v},
    \label{eq:jae}
\end{equation}
where 
\begin{equation}
    \tilde{\mu} \equiv \mu_e\frac{1}{1 + N_h/N_e}+ \mu_h\frac{1}{1 + N_e/N_h}
    \label{eq:mutilde}
\end{equation}
and
\begin{equation}
    \tilde{\Gamma} \equiv K_{\rm eff}^2 k \frac{\sigma_{m}}{\sigma_{\square}}.
    \label{eq:gammatilde}
\end{equation}
Here, $\mu_e$ and $\mu_h$ denote the mobilities of electrons and holes, respectively, while $N_e$ and $N_h$ represent their corresponding densities of states. In the derivation, Einstein’s relation $\sigma_{n,p} = e^2 N_{n,p} D_{n,p}$ was employed, where $\sigma_n$ and $D_n$ ($\sigma_p$ and $D_p$) are the conductivity and diffusion coefficient of electrons (holes), respectively. The quantity $I_{\rm SAW}$ denotes the SAW intensity per unit length, $k$ is the SAW wave number, and $K_{\rm eff}^2$ is the electromechanical coupling coefficient. The sheet conductance of the film is given by $\sigma_{\square} = \sigma_0 d$, where $\sigma_0$ and $d$ are the total conductivity and the film thickness, respectively. The characteristic conductance $\sigma_m$ is defined as $\sigma_m = v \epsilon_{\rm p}$, where $\epsilon_{\rm p}$ is the effective dielectric constant of the piezoelectric substrate~\cite{datta1986,kawada2025jap}. To derive Eqs.~(\ref{eq:jae})–(\ref{eq:gammatilde}), we assume that charge conservation holds independently for electrons and holes.

Equation~(\ref{eq:jae}) fails to account for the observed polarity reversal of the AE current between NbSe$_3$($X+90^{\circ}$) and NbSe$_3$($X$), because the sign of each component in Eq.~(\ref{eq:jae}) is independent of the SAW propagation direction. Furthermore, the magnitude of the AE current in Eq.~(\ref{eq:jae}) is significantly smaller than the experimentally observed value. As an illustrative example, we estimate the AE current within the conventional framework for NbSe$_3$($X+90^{\circ}$) at 50 K. We adopt $\mu_e = -1500\:\mathrm{cm^2/(V\cdot s)}$ and $\mu_h = 1900\:\mathrm{cm^2/(V\cdot s)}$ from a previous study~\cite{ong1978prb}, and assume $N_h/N_e = 1$ for simplicity. The parameters $K_{\rm eff}^2 = 0.011$ and $\epsilon_{\rm p} = 320\ \mathrm{pF/m}$ are used for the SAW propagation along the $X+90^\circ$ direction~\cite{kawada2025jap}. From the experimental data, we obtain $\sigma_0 = 2.7 \times 10^6\:\mathrm{(\Omega \cdot m)^{-1}}$ and $v = 3690\ \mathrm{m/s}$. $I_{\rm SAW} = 5.9\ \mathrm{W/m}$ was estimated from $P_{\rm SAW} = 0.82$ mW and a SAW aperture of 140 $\upmu$m. By substituting these parameters into Eq.~(\ref{eq:jae}) and multiplying by the width of the NbSe$_3$ flake (0.79 $\upmu$m), we obtain an AE current of approximately 1.3 pA/mW which is about four orders of magnitude smaller than the experimentally observed value. This discrepancy indicates that the contribution from the conventional model is negligibly small under the present experimental conditions.

\section{Details of the numerical calculation\label{sec:calc}}
\begin{figure}[b]
	\begin{minipage}{1.0\hsize}
		\centering
		\includegraphics[scale=0.09]{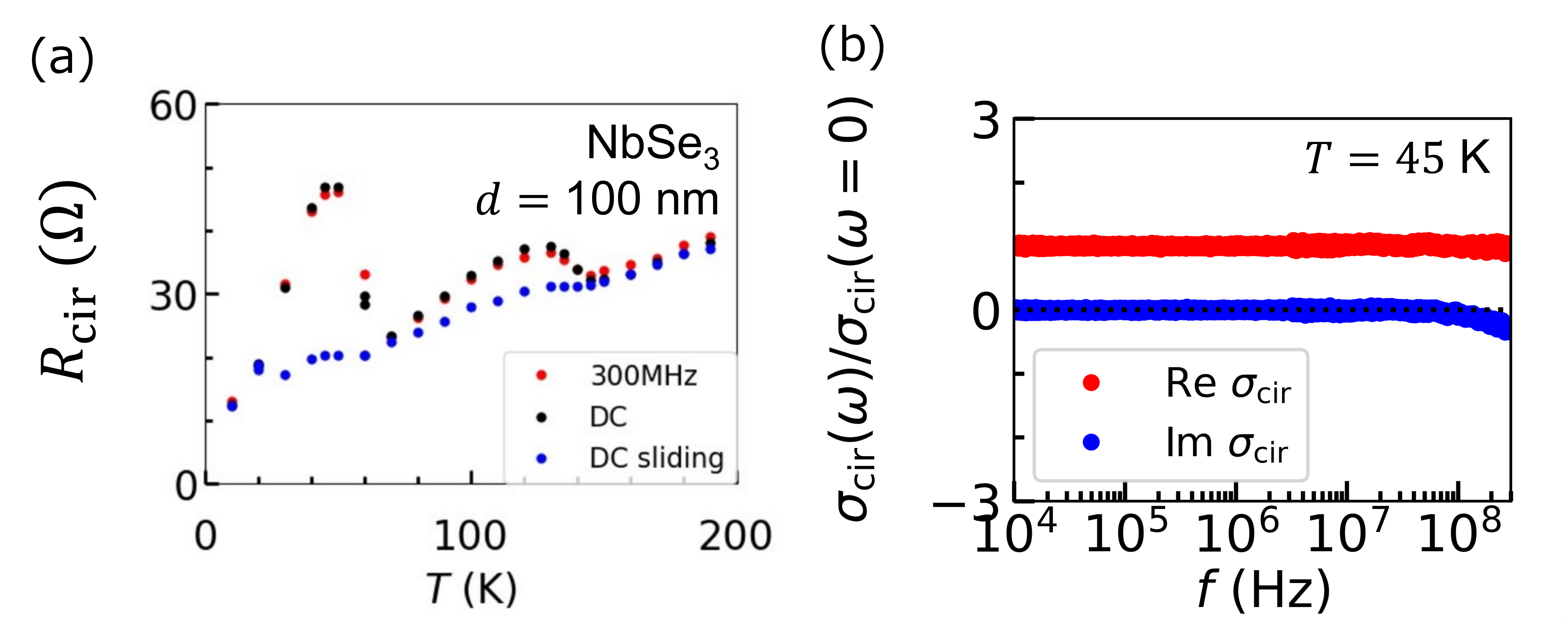}
	\end{minipage}
	\caption{(a) Temperature dependence of dc and ac circuit resistance ($R_{\rm cir}$) of 100 nm-thick NbSe$_3$ on a Si/SiO$_x$. Red, black, and blue dots represent the data with respect to the rf signal of 300 MHz, the dc voltage without CDW sliding, and the dc voltage with CDW sliding, respectively. (b) Rf frequency dependence of complex ac circuit conductivity ($\sigma_{\rm cir}(\omega)$) normalized by the dc circuit conductivity ($\sigma_{\rm cir}(\omega=0)$). Red and blue dots correspond to the real and imaginary parts, respectively. The data were obtained at $T=45$ K. \label{fig:acmeas}
	}
\end{figure}

\begin{figure}[bth]
	\begin{minipage}{1.0\hsize}
		\centering
		\includegraphics[scale=0.16]{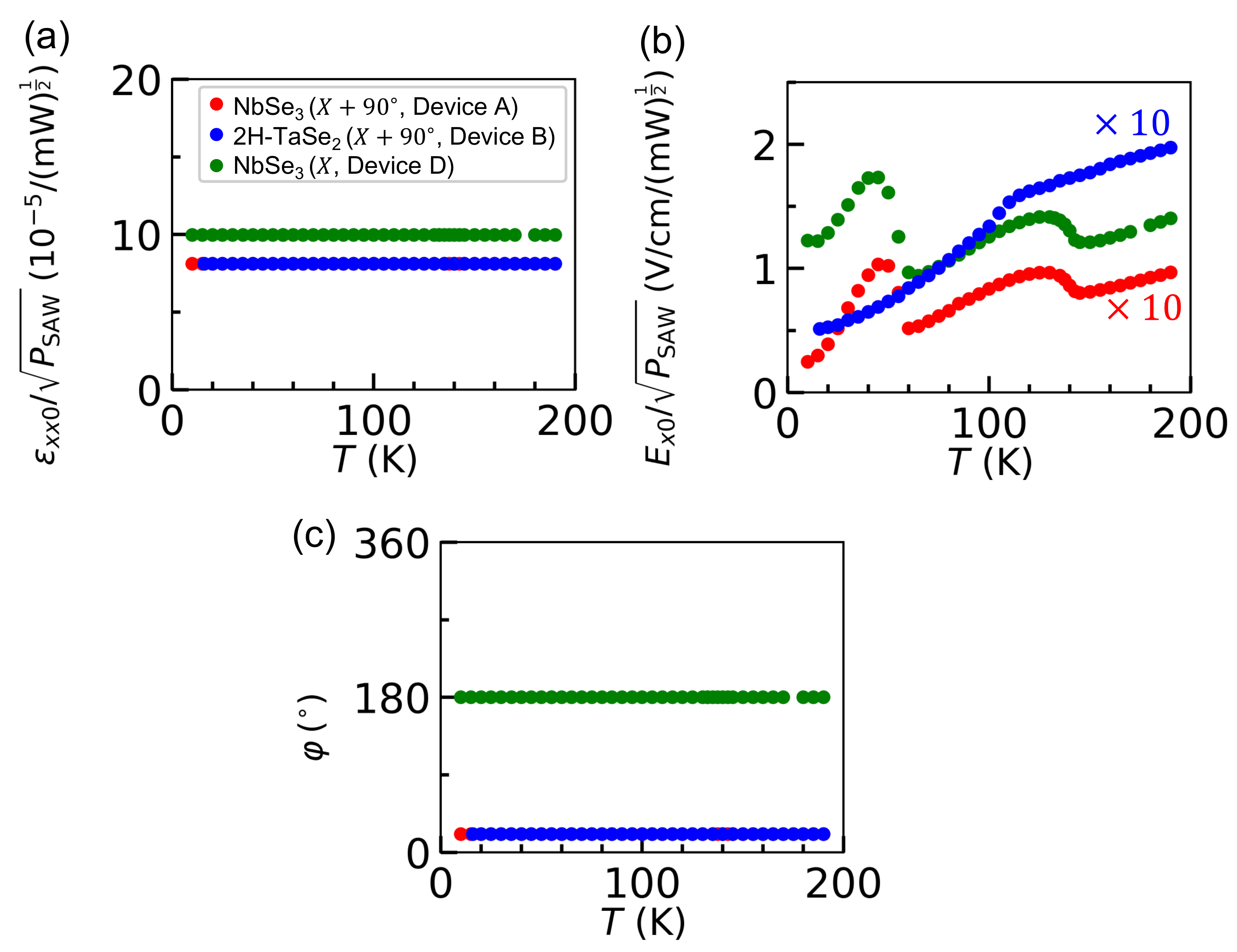}
	\end{minipage}
	\caption{Estimations of (a)~$\varepsilon_{xx0}$, (b)~$E_{x0}$, and (c)~$\varphi$. Red, blue, and green dots show results for NbSe$_3$($X+90^\circ)$, 2H-TaSe$_2$($X+90^\circ)$, and NbSe$_3$($X)$, respectively. $E_{x0}$ for NbSe$_3$($X+90^\circ)$ and 2H-TaSe$_2$($X+90^\circ)$ are plotted with a factor of ten magnification. $\varepsilon_{xx0}$ and $\varphi$ for NbSe$_3$($X+90^\circ)$ and 2H-TaSe$_2$($X+90^\circ)$ are almost equal to each other. The diffusion coefficient $D_e$ was set to 10 $\mathrm{cm^2/s}$.
		\label{fig:calc}
	}
\end{figure}

To estimate the amplitude and phase of the SAW-related quantities, we calculated the propagation characteristics of SAW in the presence of a metallic film using established theoretical frameworks~\cite{campbell1968ieee,ingebrigtsen1969,ingebrigtsen1970,datta1986,kushibiki1999ieee,kawada2025jap}. In these calculations, the electrical conductivity and relative permittivity of the film are required. Accordingly, we first measured the ac resistance of a NbSe$_3$ thin film.

A through-type coplanar waveguide incorporating a 100 nm-thick NbSe$_3$ flake was fabricated on a Si/SiO$x$ substrate, with the geometry designed to achieve a characteristic impedance of 50~$\Omega$. The $S$-parameters were measured in a two-port configuration using a VNA, with calibration performed at room temperature using the short–open–load–through (SOLT) method. The complex ac conductivity of the entire device ($\sigma_{\rm cir}$) was then extracted from the measured $S$-parameters following the procedure described in Ref.~\cite{Awan2016}.

Figure~\ref{fig:acmeas}(a) shows the $T$ dependence of the circuit resistance ($R_{\rm cir}$). The ac resistance measured at 300 MHz is found to be nearly identical to the dc resistance and does not coincide with the differential resistance obtained under CDW sliding conditions. Figure~\ref{fig:acmeas}(b) presents the frequency dependence of the complex ac conductivity at $45$~K, which is normalized by the dc conductivity. In contrast to previous reports~\cite{gruner1980}, the real part of the ac conductivity remains approximately equal to the dc value, and the imaginary part is nearly negligible. It should be noted that the slight decrease in the imaginary component at higher frequencies is attributed to insufficient calibration at low temperatures. The discrepancy between the present results and earlier studies can be ascribed to the strong pinning of the CDW in thin films. Indeed, the threshold electric field in NbSe$_3$ thin films can be up to two orders of magnitude larger than that in bulk samples~\cite{fujiwara2021}. As a result, CDW electrons are unlikely to be driven at sub-GHz frequencies.

The calculation was performed according to the methodology established in previous studies~\cite{campbell1968ieee,ingebrigtsen1969,ingebrigtsen1970,datta1986,kushibiki1999ieee,kawada2025jap}. We employ the experimentally obtained dc resistivity, a relative permittivity of $\epsilon_r = 1$, which was justified by the ac resistance measurements, and the literature values of LiNbO$_3$ parameters~\cite{kushibiki1999ieee}. For simplicity, the electronic transport in NbSe$_3$ is approximated within a single-carrier framework. The $T$ dependence of the material parameters of LiNbO$_3$ is neglected, as their variation is at most a few percent~\cite{tarumi2012low}. For SAW propagation along the $X+90^\circ$ ($X$) direction, the following parameters were employed: an electromechanical coupling coefficient $K_{\mathrm{eff}}^2 = 0.011$ (0.054)~\cite{kawada2025jap}, an effective dielectric constant $\epsilon_{\rm p} = 320\ \mathrm{pF/m}$ ($350\ \mathrm{pF/m}$)~\cite{kawada2025jap}, and a SAW aperture $W_a = 140~\upmu$m. The electron diffusion coefficient $D_e=10\ \mathrm{cm}^2/\mathrm{s}$ is assumed. We confirmed that $\varphi$, $E_{x0}$, and $\varepsilon_{xx0}$ vary by at most 2 \% over the range $D_e = 1$–$1000\ \mathrm{cm}^2/\mathrm{s}$, which includes typical values for metals~\cite{niimi2013}.

The numerical calculations for $\varepsilon_{xx0}$, $E_{x0}$, and $\varphi$ are presented in Fig.~\ref{fig:calc}, where $\varepsilon_{xx0}$ and $E_{x0}$ are normalized by the SAW power. The temperature dependence is indirectly incorporated through that of the resistivity. We found that $E_{x0}$ scales with $\rho$, whereas the other parameters does not strongly depend on $\rho$. Thus, the quantity $J_0 \cos \varphi$ is little dependent on $T$, as explicitly demonstrated in Fig.~\ref{fig:iaetase2}(a).

\bibliography{ref_NbSe3}

\end{document}